\documentclass[conference]{IEEEtran}[10pt]
\IEEEoverridecommandlockouts
\usepackage{cite}
\usepackage{amsmath,graphicx}
\usepackage[table,xcdraw]{xcolor}
\usepackage{amsmath,amssymb}
\usepackage{multirow}
\usepackage{graphicx}
\usepackage{epstopdf}
\usepackage{hyperref}
\usepackage{subcaption}
\usepackage[english]{babel}
\usepackage{enumitem}
\usepackage{breqn}
\usepackage{makecell}
\usepackage{eqparbox}
\usepackage{booktabs}
\usepackage{multirow}
\DeclareMathOperator{\E}{\mathbb{E}}
\newdimen{\algindent}
\usepackage{eso-pic}
\usepackage[linesnumbered,ruled,vlined]{algorithm2e}
\usepackage{bm}
\usepackage{comment}
\hypersetup{
    colorlinks,
    linkcolor={black!50!black},
    citecolor={blue!50!black},
    urlcolor={blue!80!black}
}
\def\BibTeX{{\rm B\kern-.05em{\sc i\kern-.025em b}\kern-.08em
    T\kern-.1667em\lower.7ex\hbox{E}\kern-.125emX}}

\begin{document}
\title{Speech enhancement deep-learning architecture for efficient edge processing}
\author{\begin{tabular}{c}Monisankha Pal, Arvind Ramanathan, Ted Wada, Ashutosh Pandey
        \end{tabular} \\
\IEEEauthorblockA{\textit{Infineon Technologies, Irvine, USA}}}

\maketitle

\begin{abstract}
Deep learning has become a \emph{de facto} method of choice for speech enhancement tasks 
with significant improvements in speech quality. However, real-time processing with reduced size and computations for low-power edge devices drastically degrades speech quality. Recently, transformer-based architectures have greatly reduced the memory requirements and provided ways to improve the model performance through local and global contexts. However, the transformer operations remain computationally heavy. In this work, we introduce WaveUNet squeeze-excitation Res2 (WSR)-based metric generative adversarial network (WSR-MGAN) architecture that can be efficiently implemented on low-power edge devices for noise suppression tasks while maintaining speech quality. We utilize multi-scale features using Res2Net blocks that can be related to spectral content used in speech-processing tasks. In the generator, we integrate squeeze-excitation blocks (SEB) with multi-scale features for maintaining local and global contexts along with gated recurrent units (GRUs). The proposed approach is optimized through a combined loss function calculated over raw waveform, multi-resolution magnitude spectrogram, and objective metrics using a metric discriminator. Experimental results in terms of various objective metrics on VoiceBank+DEMAND and DNS-2020 challenge data sets demonstrate that the proposed speech enhancement (SE) approach outperforms the baselines and achieves state-of-the-art (SOTA) performance in the time domain.
\end{abstract}

\begin{IEEEkeywords}
Edge ML, metric discriminator, Res2Net, speech enhancement, time domain. 
\end{IEEEkeywords}

\section{Introduction}
Speech enhancement (SE) \cite{loizou2007speech} is the task of improving the perceptual quality and intelligibility of speech signals contaminated by additive background noise. It is an indispensable component in speech recognition, hearing aids, and modern smart speakers. In recent years, the evolution of deep neural networks (DNNs) and the plethora of DNN-based speech-denoising works confirm their superiority over traditional signal processing based methods in low signal-to-noise-ratio (SNR) conditions \cite{xu2013experimental} with non-stationary noises such as dogs barking, birds chirping, keyboard typing, \emph{etc}. 
\par
Speech enhancement systems can be categorized into time-frequency (TF) and time (T) domain-based systems. In the TF paradigm, the estimation of certain masks, e.g., ideal ratio mask \cite{wang2021neural}, spectral magnitude mask \cite{fu2019metricgan, fu2021metricgan+}, and usage of spectral features \cite{hao2021fullsubnet, isik2020poconet, fu2021metricgan+} was investigated by utilizing noisy magnitude spectrogram while ignoring the phase \cite{valin2018hybrid, abdulatif2021investigating, fu2021metricgan+}. To mitigate the challenging phase issue, the approaches of incorporating the complex spectrogram and predicting either the complex ratio mask \cite{hu2020dccrn} or directly mapping to the real and imaginary components of clean speech \cite{tan2019complex} have been proposed with varying degrees of success. 
\par
On the other hand, the \emph{time-domain} SE methods directly predict the clean speech from the noisy speech input without considering any hand-crafted spectral features. These compelling end-to-end (E2E) approaches enhance the raw noisy speech with no explicit assumption about the data and thus avoid the transformation overheads. To capture both sequential and local patterns, WaveCRN, which is an efficient E2E convolutional recurrent neural network based encoder-decoder (CED) architecture, was introduced \cite{hsieh2020wavecrn}. Insertion of a temporal convolutional module between the CED with skip connections to learn long-range dependencies from the past was explored \cite{pandey2019tcnn}. Recently, architectures exploiting CED with LSTMs \cite{defossez2020real}, multi-head self-attention (MHSA) \cite{wang2021tstnn, kong2022speech}, and conformer layers \cite{kim2021se} have been proposed and are leading the performance benchmark in terms of speech quality. 
Although the E2E methods remain the best-performing methods, they are computationally heavy and not suitable for edge ML applications. In an era where privacy and sustainable living are paramount, the \emph{edge ML} is growing rapidly in popularity for consumer devices such as hearing aids and smart speakers \cite{fedorov2020tinylstms, shankar2020real, wang2021noisy}. 
It is imperative to bring the SOTA methods to the edge ML space in power efficient manner.  

\begin{figure*}[!t]
 \centering
  \includegraphics[width=\textwidth]{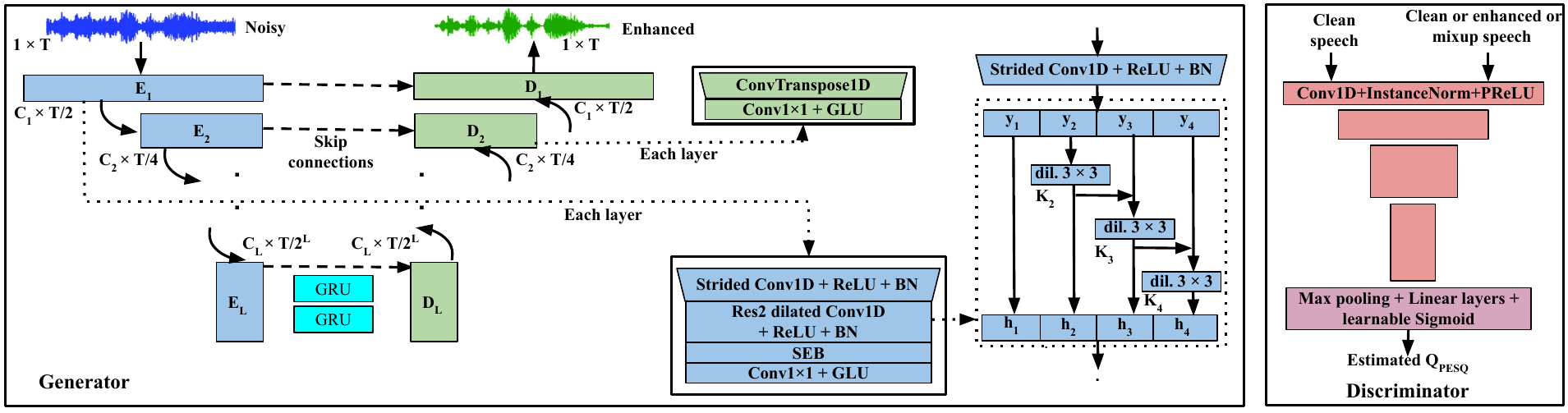}
  \caption{An overview of the proposed WSR-MGAN architecture.}
  \label{fig1}
\end{figure*}

\par
In this paper, we present an edge-efficient WaveUNet squeeze-excitation Res2 (WSR)-based metric generative adversarial network (\emph{WSR-MGAN}) architecture in time domain, inspired by Res2Net's promising performances on computer vision \cite{gao2019res2net}, speech anti-spoofing \cite{li2021replay}, and speaker verification \cite{desplanques2020ecapa} tasks. Specifically, the proposed \emph{U-Net-based} time-domain SE incorporates \emph{Res2Net architecture} and \emph{squeeze-excitation block (SEB)} \cite{hu2018squeeze} inside each convolution encoder layer of the WSR-MGAN network, where the Res2Net block enables efficient extraction of multi-scale features at granular levels and increases the receptive fields, and SEB assists in modeling the channel inter-dependencies by using adaptive channel weights. Furthermore, we employ two uni-directional gated recurrent unit (GRU) layers in the bottleneck to learn the long-range dependencies, and introduce a \emph{metric discriminator} to estimate the perceptual evaluation of speech quality (PESQ) score and optimize the generator with respect to the PESQ-related loss function to further improve the speech quality. The main contributions of this work are (a) the improved lite U-Net architecture with reduced number of parameters and complexity and (b) the usage of a metric discriminator in the time domain via linear mixup strategy between the enhanced and the clean speech signals to stabilize the GAN training for increased SE performance.

\section{Edge Efficient Deep Speech Enhancement}
\label{section3}
\subsection{Generator}
The overview of the generator utilizing the U-Net structure \cite{ronneberger2015u} is shown in Figure \ref{fig1}. The CED architecture is combined with skip connections between them and two GRU layers as the bottleneck. Inside each encoder layer, we efficiently integrate Res2Net and SEB to capture local context and extract speech-related multi-scale features, and to assist in modeling channel inter-dependencies using learnable channel weights. Uni-directional GRUs are chosen, as they capture temporal context well and can be implemented efficiently on edge devices. The detailed design components are described in the following subsections. 

\subsubsection{Encoder}
Given the input raw noisy waveform $\mathbf{y} \in \mathbb{R}^{B \times C \times T}$, the encoder ($E$) which comprises of $L$ layers outputs a latent representation $E(\mathbf{y}) = \mathbf{z}$. Here, $B$, $C$, and $T$ are batch size, no of channels (1 for input raw noisy waveform), and length of input time-domain speech data respectively. The initial layer inside each encoder layer is composed of a strided 1-D convolution (conv1D) that downsamples the input in length and increases the number of channels. Each encoder layer then contains a Res2Net with SEB followed by a Conv1$\times$1 layer with a gated linear unit (GLU) activation function. Following the conv1D layer inside Res2Net, a rectified linear unit (ReLU) activation function and a batch normalization (BN) layer are applied. Every $i$-th encoder layer outputs $\textnormal{min}(2^{i-1}H, C_m)$ channels, where $H$ represents the number of hidden channels of the first encoder layer and $C_m$ represents the maximum allowed channel dimension. Inside the encoder all the 1-D convolutions are causal.

\paragraph{Res2Net and SEB} 
The Res2Net \cite{gao2019res2net} architecture with stronger multi-scale feature extraction ability increases the number of available receptive fields. This is attained by using a smaller group of 3$\times$3 convolution filters within one block. These different filter groups are connected in a hierarchical residual style. After the strided conv1D layer, we  split the feature maps $(\mathbf{y})$ by the channel dimension $C$ into $s$ subsets, each denoted by $\mathbf{y}_i$, where $i \in \{1, 2, \ldots , s\}$. Except $\mathbf{y}_{1}$, all other $\mathbf{y}_i$ is processed through a 3$\times$3 convolution filter, denoted by $\mathbf{K}_i()$. The output from the Res2Net block is obtained as
\begin{equation}
\footnotesize
    \mathbf{h}_{i} = \begin{cases}
    \mathbf{y}_{i} & i = 1\\
    \mathbf{K}_i(\mathbf{y}_{i}) & i = 2\\
    \mathbf{K}_i(\mathbf{y}_{i} + \mathbf{h}_{i-1}) & 2 < i  \le s.
  \end{cases}
\end{equation}
where $s$ represents the scale dimension. The output of the Res2Net block represents a different number and combination of receptive fields/scales. Furthermore, to increase the context information, we use dilated convolution with dilation factor 2 in addition to ReLU and BN inside the Res2Net block. 

SEB \cite{hu2018squeeze} adaptively re-calibrates channel-wise feature responses by modeling global channel inter-dependencies at almost zero computational cost. The squeeze component of SEB reduces each channel to a single numeric value by global average pooling across the time dimension. 
The descriptors 
are then passed through the excitation module to calculate weights for each channel.
The resulting vector containing channel weights is used to perform channel-wise multiplication. The SEB output is fed to each Conv1$\times$1 layer that doubles the number of channels and then GLU halves the number of channels to bring back the same. 

\subsubsection{Bottleneck}
Next in the bottleneck, a sequence modeling GRU network $G$ takes the latent outputs $\mathbf{z}$ from the encoder and outputs a non-linear transformation of the same size, i.e., $\hat{\mathbf{z}} = G(\mathbf{z})$. The GRU network has two uni-directional GRU layers
that slightly outperform LSTMs for SE tasks.

\subsubsection{Decoder}
Finally, the decoder network $D$ takes bottleneck layers output and produces an estimation of the clean speech signal, i.e., $\hat{\mathbf{x}} = D(\hat{\mathbf{z}})$. Each decoder layer is composed of a Conv1$\times$1 followed by GLU activation and then a transposed 1D convolution (ConvTranspose1D). Moreover, a skip connection connects each decoder layer with the corresponding encoder layer. The ConvTranspose1D in each decoder layer is causal and performs the reverse operation of conv1D in the encoder layer.

\subsection{Metric discriminator}
The loss functions used to optimize the SE model are not fully correlated with the objective evaluation metrics that consider human auditory perception. Moreover, we cannot directly employ PESQ evaluation as a loss function because it is not differentiable. To overcome this gap, we deploy a metric discriminator to mimic the target evaluation metrics and optimize it using adversarial losses. The discriminator network shown in Figure \ref{fig1} comprises four conv blocks with a conv1D layer, followed by instance normalization and PReLU activation function. Afterward, adaptive max pooling is followed by two linear layers, and learnable Sigmoid \cite{fu2021metricgan+} activation function is applied at the end.
The discriminator takes two input speech signals as clean-clean, clean-enhanced, or clean-mixup \cite{zhang2018mixup}, and it is trained to predict the corresponding normalized PESQ scores that represent the ground-truth labels.

\subsection{Loss function}
The adversarial training in WSR-CMGAN follows a min-min optimization task over the generator $(\mathcal{G})$ and the discriminator $(\mathcal{D})$. In addition to the generator adversarial loss, we incorporate L1-distance loss over the waveform and multi-resolution STFT (MRSTFT) loss  \cite{defossez2020real} over the magnitude spectrogram. 
Let $S(\mathbf{x}; \boldsymbol{\phi}) = |STFT(\mathbf{x})|$ be the linear-scale magnitude spectrogram of clean speech $\mathbf{x}$, where $\boldsymbol{\phi}$ represents the STFT hyper-parameters such as number of FFT bins, hop sizes, and window lengths. We employ full-band MRSTFT loss for our model training as:
\begin{equation}
\footnotesize 
\mathcal{L}_{\textnormal{mrstft}}(\mathbf{x}, \hat{\mathbf{x}}) = \sum_{i=1}^M \frac{||S(\mathbf{x}; \boldsymbol{\phi}_i) - S(\hat{\mathbf{x}}; \boldsymbol{\phi}_i)||_F}{||S(\mathbf{x}; \boldsymbol{\phi}_i)||_F} + \frac{1}{T} ||\textnormal{log} \frac{S(\mathbf{x}; \boldsymbol{\phi}_i)}{S(\hat{\mathbf{x}}; \boldsymbol{\phi}_i)}||_1
\end{equation}
where $||\cdot||_F$ and $||\cdot||_1$ are the Frobenius norm and L1 norm respectively, $M$ denotes the total number of resolutions, $T$ denotes the length of the input noisy speech signal, and $ \boldsymbol{\phi}_i$ represents STFT parameters at $i$-th resolution. Finally, our loss function to train the generator model is given by
\begin{equation}
\footnotesize 
    \mathcal{L}_{\mathcal{G}} = {\alpha}_1 ||\mathbf{x} - \hat{\mathbf{x}}||_1 + {\alpha}_2 \mathcal{L}_{\textnormal{mrstft}}(\mathbf{x}, \hat{\mathbf{x}}) + {\alpha}_3  \E_{\mathbf{x}, \hat{\mathbf{x}}} \left[||\mathcal{D}(\mathbf{x}, \hat{\mathbf{x}}) - 1||^2 \right] \label{loss function}
\end{equation}
where ${\alpha}_1$, ${\alpha}_2$, and ${\alpha}_3$ are the weights of three different loss functions. The first term represents the L1-distance loss and the last term represents the generator adversarial loss.
\par
Motivated by the success of the mixup augmentation \cite{zhang2018mixup} strategy to stabilize GAN training, we employ a mixup between clean and enhanced speech signals. It helps regularize the gradients of the discriminator and ensures a stable source of gradient information for the generator. We use the following mixup:
\begin{equation}
\footnotesize 
\hat{\mathbf{x}}_{\textnormal{mix}} = \lambda x + (1-\lambda)\hat{\mathbf{x}}
\end{equation}
where $\lambda \in [0, 1]$ and sampled from a Beta distribution.
Finally, the discriminator loss can be expressed as
\footnotesize 
\begin{align}
 \mathcal{L_D} = \E_{\mathbf{x}, \mathbf{x}} & \left[||\mathcal{D}(\mathbf{x}, \mathbf{x}) - 1||^2 \right] + \E_{\mathbf{x}, \hat{\mathbf{x}}} \left[||\mathcal{D}(\mathbf{x}, \hat{\mathbf{x}}) - Q_{\textnormal{PESQ}}(\mathbf{x}, \hat{\mathbf{x}})||^2 \right] \notag\\
  & + \E_{\mathbf{x}, \hat{\mathbf{x}}_{\textnormal{mix}}} \left[||\mathcal{D}(\mathbf{x}, \hat{\mathbf{x}}_{\textnormal{mix}}) - Q_{\textnormal{PESQ}}(\mathbf{x}, \hat{\mathbf{x}}_{\textnormal{mix}})||^2 \right]
\end{align}
\normalsize
where $Q_{\textnormal{PESQ}}(\mathbf{x}, \hat{\mathbf{x}})$ and $Q_{\textnormal{PESQ}}(\mathbf{x}, \hat{\mathbf{x}}_{\textnormal{mix}})$ refers to the normalized true PESQ scores between $(\mathbf{x}, \hat{\mathbf{x}})$ and $(\mathbf{x}, \hat{\mathbf{x}}_{\textnormal{mix}})$ respectively.

\begin{table*}[!t]
\footnotesize
\caption{SE performance on VoiceBank+DEMAND test set. Here $``-"$ denotes the result is not provided in the original paper, $``\textnormal{Impl.}"$ means the system is implemented and $``\textnormal{Base}"$ means the base proposed model. The results of models with $<$ 10M parameters are highlighted in grey. The results after the double horizontal line are from different proposed methods.}
\centering
\setlength{\tabcolsep}{6.5pt}
\begin{tabular}{ccccccccccc}
\Xhline{2.5\arrayrulewidth}
Method                                      & Domain         & $\#$par.(M) & MACS(G) & RTF  & WB-PESQ & NB-PESQ & CSIG & CBAK & COVL & STOI \\ \Xhline{2.5\arrayrulewidth}
Noisy                                       & --             & --            & --      & --   & 1.97    & 2.55    & 3.34 & 2.45 & 2.63 & 0.92 \\
\rowcolor[HTML]{C0C0C0} 
RNNoise (Impl.) \cite{valin2018hybrid}              & TF & \textbf{0.09} & \textbf{0.04}    & \textbf{0.03} & 2.48    & 3.24    & 3.64 & 2.92 & 3.03 & 0.93 \\
\rowcolor[HTML]{C0C0C0} 
DCCRN \cite{hu2020dccrn}                        & TF & 3.70          & 14.36   & 2.19 & 2.84    & --      & 4.03 & 2.97 & 3.43 & 0.94 \\
\rowcolor[HTML]{C0C0C0}
MetricGAN+ \cite{fu2021metricgan+}          & TF & 2.60          & --      & --   & 3.15    & --      & 4.14 & 3.16 & 3.64 & 0.93   \\
\rowcolor[HTML]{C0C0C0} 
TSTNN \cite{wang2021tstnn}                 & T       & 0.92           & 19.95   & 2.40 & 2.96    & --      & 4.33 & 3.53 & 3.67 & 0.95 \\
\rowcolor[HTML]{C0C0C0} 
CleanUNet-lite (Impl.) \cite{kong2022speech}     & T       & 2.19           & 1.90    & 0.12 & 2.74    & 3.38    & 4.20 & 3.34 & 3.48 & 0.94 \\
Wave-U-Net \cite{macartney2018improved}   & T       & --    & --  & --         & 2.40    & --      & 3.52 & 3.24 & 2.96 & --   \\
FAIR denoiser \cite{defossez2020real}      & T       & 33.53          & --      & --   & 2.84    & 3.33    & 4.27 & 3.38 & 3.57 & 0.95 \\
CleanUNet \cite{kong2022speech}            & T       & 46.07          & --      & --   & 2.91    & 3.41    & 4.34 & 3.42 & 3.65 & \textbf{0.96} \\
SE-Conformer \cite{kim2021se}              & T       & --             & --      & --   & 3.13    & --      & 4.45 & 3.55 & 3.82 & 0.95 \\ \hline \hline
\rowcolor[HTML]{C0C0C0}  WaveUNet (Base)                              & T       & 1.33           & 
1.85    & 0.08 & 2.68    & 3.34    & 4.14 & 3.29 & 3.42 & 0.94 \\
\rowcolor[HTML]{C0C0C0} + GRU                            & T       & 1.53           & 
1.86 & 0.10 & 2.87    & 3.49    & 4.29 & 3.40 & 3.59 & 0.95 \\
\rowcolor[HTML]{C0C0C0} + GRU + Res2                              & T       & 1.60           & 
1.96 & 0.13 & 2.92    & 3.55    & 4.32 & 3.38 & 3.63 & 0.95 \\
\rowcolor[HTML]{C0C0C0} + GRU + Res2 + SEB                              & T       & 1.62           & 
1.96 & 0.14 & 2.98    & 3.57    & 4.40 & 3.50 & 3.71 & 0.95 \\
\rowcolor[HTML]{C0C0C0} + GRU + Res2 + SEB, with Disc.                              & T       & 1.62           & 
1.96 & 0.14 & 3.06    & 3.66    & 4.41 & 3.50 & 3.75 & 0.95 \\
\rowcolor[HTML]{C0C0C0} 
\begin{tabular}[c]{@{}c@{}}+ GRU + Res2 + SEB, with Disc.\\ + mixup = WSR-MGAN-lite\end{tabular} & T & 1.62       & 
1.96 & 0.14 & 3.09    & 3.67    & 4.46 & 3.55 & 3.80 & 0.95 \\
WSR-MGAN                                   & T       & 38.50          & 
13.49 & 0.28   & \textbf{3.18}    & \textbf{3.76}    & \textbf{4.50} & \textbf{3.63} & \textbf{3.88} & 0.95 \\  \Xhline{2.5\arrayrulewidth}
\end{tabular} \label{table1}
\end{table*}

\section{Experimental setup}
\label{section4}
\subsection{Dataset}
The VoiceBank+DEMAND \cite{valentini2016investigating} is a publicly available standard database for SE works. The training set contains 11,572 utterances (total time $\thicksim$10 h) from 28 speakers and the test set consists of 824 utterances from two unseen speakers. In the training set, the clean utterances are mixed with background noise at SNRs of $[0, 5, 10, 15]$ dB with $8$ different noise types taken from the DEMAND database and two artificial noise types. In the test set, the clean utterances are mixed with 5 unseen noise types from DEMAND at SNRs of 2.5 dB, 7.5 dB, 12.5 dB and 17.5 dB. 
\par
DNS-2020 challenge database \cite{reddy2020interspeech} contains clips from more than 10K speakers with 16 kHz sampling frequency. The clean set has more than 500 hours of clean speeches, and the noise set is over 180 hours from 150 noise classes. We generate 500 hours of clean and noisy speech pairs as the training set with 31 SNR levels (-5 to 25 dB). We adopt no-reverb synthetic test set for our and other system's objective evaluation.
\subsection{Implementation details}
We randomly crop $1.5$-sec and $10$-sec clips for each training data pair $(\mathbf{y}, \mathbf{x})$ for VoiceBank+DEMAND and DNS-2020 experiments, respectively. For VoiceBank+DEMAND experiments, we apply Remix and BandMask \cite{defossez2020real, kong2022speech} augmentation, and for DNS-2020 no-reverb experiments, we apply RevEcho \cite{defossez2020real, kong2022speech} augmentation. We use $L = 8$ and each encoder-decoder has $H = 64$, stride 2, and kernel size $4$. The maximum allowed channel dimension $C_m$ is set to $128$ and $768$ for the lite model and heavy model, respectively. 
In the Res2Net layers, the group of convolution filters is $3\times3$, the scale dimension is $4$, and the dilation factor is set to $2$. Two uni-directional GRU layers with $C_m$ hidden units are employed as the bottleneck layers. 
\par
We train our proposed model using Adam optimizer with momentum $\beta_1 = 0.9$ and $\beta_2 = 0.999$. A linear warmup with cosine annealing learning rate with a maximum value of $2e-4$ is used. We train our proposed setup for $250$K iterations with batch size $8$ on two GPUs for all the experiments.

\section{Experimental results}
\label{section5}
In this section, we evaluate and compare our proposed SE system against several baselines and recent SOTAs. We chose a pair of commonly used objective metrics to evaluate the SE performance: PESQ score \cite{recommendation2001perceptual} and short-time objective intelligibility (STOI) score \cite{taal2011algorithm}. In addition, we employ three mean opinion score prediction-based metrics: speech signal distortion (CSIG), intrusiveness of background noise (CBAK), and overall quality (COVL) \cite{loizuquality}. The higher the values of all of the above-mentioned metrics are, the better the performance is. Our proposed setup is similar to CleanUNet \cite{kong2022speech}, and both are time-domain based. Therefore, CleanUNet-based SE is our main baseline system.

\subsection{Results on VoiceBank+DEMAND dataset}
 Table \ref{table1} summarizes all the results on the VoiceBank+DEMAND test set. All the models are trained using the original VoiceBank+DEMAND training set except RNNoise. We train an RNNoise \cite{valin2018hybrid} model based on the VoiceBank+DEMAND dataset that we have prepared similarly to the original training set.  We implement CleanUNet-lite, a lighter model with one self-attention layer, and the maximum channel dimension $C_m$ is set to 128. We observe from Table \ref{table1} that the proposed time-domain WSR-MGAN-lite approach with $1.62$M model parameters yields better results as compared to both heavier FAIR denoiser and CleanUNet models as well as other lighter models in the table. Moreover, the WSR-MGAN-lite system achieves significant improvements in CSIG, CBAK and COVL scores over the metric discriminator based approach MetricGAN+. Our framework also outperforms transformer-based time-domain methods, such as TSTNN and CleanUNet. Finally, the proposed heavier model WSR-MGAN attains improvements in all evaluation metrics over the FAIR denoiser, CleanUNet, and SE-Conformer, and achieves SOTA performance in time domain.

\subsection{Model footprints and computational complexity}
We compare and report the model footprints and computational complexity between the SE models in Table \ref{table1}. 
The model footprints are measured using the number of parameters, and the computational complexity is computed using minimal amount of multiply-accumulate operation per second (MACS). MACS is measured using `ptflop' PyTorch package\footnote{\url{https://github.com/sovrasov/flops-counter.pytorch}}, and MACS for TSTNN is computed based on the official TSTNN model code\footnote{\url{https://github.com/key2miao/TSTNN/tree/master}}. Real-time factors (RTFs) are measured based on the average inference time of five runs on a Core i-5 CPU using 1 sec of dummy data. It is observed that although the performance of TSTNN is competitive with only 0.92M model parameters, the MACS and RTF are large, hence TSTNN is not suitable for edge-device implementation. The proposed WSR-MGAN-lite model achieves the best results while requiring a minimal amount of footprints, MACS, and RTF over other lighter models as shown in Table \ref{table1}. 

\subsection{Ablation study}
We perform an ablation study on the VoiceBank+DEMAND test set to examine the contribution of each component towards speech enhancement. The results are summarized in Table \ref{table1}. First, we evaluate the SE performance of our base model named WaveUNet in the table. The base model contains only the strided Conv1D layer and then the Conv1$\times$1 layer in the encoder (no Res2, SEB, GRU, Disc.). The corresponding results are reported in row 11 of Table \ref{table1}. Thereafter, we include the GRU layers in the bottleneck of the base model which leads to significant performance improvement. By comparing the row 11 and row 12 results, we can say that the GRU layers, which capture the long-term context, are critical for the SE task. We then investigate the influence of Res2 layers in SE by including them inside each encoder layer. The results in row 13 of Table \ref{table1} indicate improvement in WB-PESQ by 0.05 with a minor increase in model footprint and complexity. We further add SEB after the Res2 layer inside each encoder layer and the corresponding results in row 14 indicate SEB is important for boosting the SE performance. Finally, we introduce the metric discriminator (Disc.) in the proposed system (row 15) which yields further improvement in WB-PESQ by 0.08 with no further increase in model footprint and complexity. In the end, we show the effect of introducing a mixup strategy (row 16) between the clean and enhanced speech and use that as the input to the discriminator. This is our final proposed system WSR-MGAN-lite. We conclude that the mixup strategy further helps in improving the SE objective metrics results slightly.

\subsection{Results on DNS-2020 dataset}
We compare the proposed system with several baselines and recent SOTA systems on the DNS-2020 challenge dataset. The experimental results on the no-reverb partition of the synthetic test set of DNS-2020 are shown in Table \ref{table2}. 
PoCoNet and DCCRN are the top-ranked methods in the non-real-time and real-time track of the DNS-2020 challenge. FullSubNet exhibits better results than both PoCoNet and DCCRN. The table shows that the proposed WSR-MGAN model achieves better objective scores as compared to all the baselines. Furthermore, WSR-MGAN yields results that are similar to those from the recent SOTA in the time domain, i.e., CleanUNet. However, the CleanUNet model comes with some extra complexity. 
The lighter version of the proposed model (WSR-MGAN-lite) with 1.62M parameters provides better SE performance than most of the time and time-frequency domain-based lighter models.  

\begin{table}[!t]
\footnotesize
\caption{SE performance evaluation on DNS-2020 no-reverb test set. Here $``-"$ denotes the result is not provided in the original paper. The results of models with $<$ 10M parameters are highlighted in grey.}
\centering
\setlength{\tabcolsep}{0.4pt}
\begin{tabular}{ccccccccc}
\Xhline{2.5\arrayrulewidth}
Method                & $\#$par.(M) & WB-PESQ & NB-PESQ & CSIG & CBAK & COVL & STOI \\
\Xhline{2.5\arrayrulewidth}
Noisy                          & --             & 1.58    & 2.16    & 3.19 & 2.53 & 2.35 & 0.92 \\
\rowcolor[HTML]{C0C0C0} RNNoise \cite{valin2018hybrid}              & \textbf{0.09}           & 2.19    & 2.77    & 3.77 & 2.94 & 2.95 & 0.94 \\
\rowcolor[HTML]{C0C0C0} FullSubNet \cite{hao2021fullsubnet}         & 5.60            & 2.78    & 3.31    & --   & --   & --   & 0.96 \\
\rowcolor[HTML]{C0C0C0} DCCRN \cite{hu2020dccrn}        & 3.70            & --      & 3.27    & --   & --   & --   & -- \\
\rowcolor[HTML]{C0C0C0} CleanUNet-lite \cite{kong2022speech}           & 2.19          & 2.51    & 3.04    & 4.13 & 3.39 & 3.33 & 0.96 \\

PoCoNet \cite{isik2020poconet}              & 50             & 2.75    & --      & 4.08 & 3.04 & 3.42 & -- \\
FAIR denoiser \cite{defossez2020real}              & 33.53          & 2.66    & 3.23    & 4.15 & 3.63 & 3.42 & \textbf{0.97} \\

CleanUNet \cite{kong2022speech}        & 46.07          & 3.15    & 3.55    & 4.62 & \textbf{3.89} & 3.93 & \textbf{0.97} \\
\Xhline{2.5\arrayrulewidth}
\rowcolor[HTML]{C0C0C0} WSR-MGAN-lite    & 1.62           & 2.73    & 3.28    & 4.26 & 3.41 & 3.51 & 0.96 \\
WSR-MGAN      & 38.50          & \textbf{3.17}    & \textbf{3.59}    & \textbf{4.63} & 3.78 & \textbf{3.95} & \textbf{0.97} \\
\Xhline{2.5\arrayrulewidth}\label{table2}
\end{tabular}
\end{table}

\section{Conclusions}
\label{section6}
We presented a novel WSR-MGAN architecture for time-domain single-channel speech enhancement. Res2Net adopted inside the encoder layers of the GAN generator enables multiple feature scales which effectively improves the SE performance, and SEB integrated with Res2Net models the channel inter-dependencies and helps improve the performance further. Furthermore, we employed a metric discriminator to optimize the enhancement model based on the PESQ-related loss function.
Collectively, architectural changes presented in this paper create edge-friendly models that utilizes a fraction of the normal compute resources.
Experimental results on two standard SE databases show that the presented model provides the SOTA performance in the time domain. A lighter model with the architecture of the proposed model provides the best-in-class noise suppression performance with $10$x less computational cost. 

\bibliography{refs}

\end{document}